\documentclass[twocolumn]{aastex62}

\usepackage{xcolor}
\DeclareRobustCommand{\ion}[2]{%
\relax\ifmmode
\ifx\testbx\f@series
{\mathbf{#1\,\mathsc{#2}}}\else
{\mathrm{#1\,\mathsc{#2}}}\fi
\else\textup{#1\,{\mdseries\textsc{#2}}}%
\fi}

\newcommand{\msun}{M$_\odot$}

\newcommand{\kms}{${\rm km \, s^{-1}}$}
\newcommand{\s}{\,{\rm s}}

\newcommand{\g}{\,{\rm g}}

\newcommand{\mpfit}{\texttt{MPFIT}}
\newcommand{\ha}{H$\alpha$}
\newcommand{\hb}{H$\beta$}

\graphicspath{{./}{figures/}}

\received{\today}
\revised{}
\accepted{}

\shorttitle{Discovering remnants with IFS}
\shortauthors{H. Mart\'{i}nez-Rodr\'{i}guez et al.}

\begin{document}

\title{Recovering lost light: discovery of supernova remnants with integral field spectroscopy}

\author[0000-0002-1919-228X]{H\'{e}ctor Mart\'{i}nez-Rodr\'{i}guez}
\affiliation{Department of Physics and Astronomy and Pittsburgh Particle Physics, Astrophysics and Cosmology Center (PITT PACC), University of Pittsburgh, 3941 O'Hara Street, Pittsburgh, PA 15260, USA}
\correspondingauthor{H\'{e}ctor Mart\'{i}nez-Rodr\'{i}guez}
\email{hector.mr@pitt.edu}
\author[0000-0002-1296-6887]{Llu\'{i}s Galbany}
\affiliation{Institute of Space Sciences (ICE, CSIC), Campus UAB, Carrer de Can Magrans, s/n, E-08193 Barcelona, Spain.}
\affiliation{Institut d’Estudis Espacials de Catalunya (IEEC), E-08034 Barcelona, Spain.}
\author[0000-0003-3494-343X]{Carles Badenes}
\affiliation{Department of Physics and Astronomy and Pittsburgh Particle Physics, Astrophysics and Cosmology Center (PITT PACC), University of Pittsburgh, 3941 O'Hara Street, Pittsburgh, PA 15260, USA}
\author[0000-0003-0227-3451]{Joseph P. Anderson}
\affiliation{European Southern Observatory, Alonso de C\'ordova 3107, Vitacura, Casilla 190001, Santiago, Chile.}
\affiliation{Millennium Institute of Astrophysics MAS, Nuncio Monsenor Sotero Sanz 100, Off.
104, Providencia, Santiago, Chile}
\author[0000-0002-3827-4731]{Inmaculada Domínguez}
\affiliation{Departamento de F\'isica Te\'orica y del Cosmos, Universidad de Granada, E-18071 Granada, Spain.}
\author[0000-0002-1132-1366]{Hanindyo Kuncarayakti}
\affiliation{Tuorla Observatory, Department of Physics and Astronomy, University of Turku, V\"ais\"al\"antie 20, FI-21500 Piikki\"o, Finland}
\author[0000-0002-3464-0642]{Joseph D. Lyman}
\affiliation{Department of Physics, University of Warwick, Coventry CV4 7AL, UK.}
\author[0000-0001-6444-9307]{Sebasti\'an F. S\'anchez}
\affiliation{Instituto de Astronom\'ia, Universidad Nacional Aut\'oonoma de M\'exico, A. P. 70-264, C.P. 04510, M\'exico, D.F., Mexico}
\author[0000-0001-7299-8373]{Jos\'e M. V\'ilchez}
\affiliation{Instituto de Astrof\'isica de Andaluc\'ia - CSIC, Glorieta de la Astronom\'ia s.n., 18008 Granada, Spain.}
\author[0000-0001-5510-2424]{Nathan Smith}
\affiliation{Steward Observatory, University of Arizona, 933 N. Cherry Ave., Tucson, AZ 85721, USA.}
\author[0000-0002-0763-3885]{Dan Milisavljevic}
\affiliation{Department of Physics and Astronomy, Purdue University, West Lafayette, IN 47907, USA}
\affiliation{Integrative Data Science Initiative, Purdue University, West Lafayette, IN 47907, USA}

\begin{abstract}

We present results from a systematic search for broad ($\geq$ 400 \kms) 
%\textcolor{red}{potser és una mica confús. Vols dir que no resolem el remnant en si (suposo), però és una mica raro la paraula perquè precisament amb IFS podem 'resoldre' espacialment aquests blobs (cosa que n podries fer amb long-slit. També està a dos altres llocs del text ('unresolved').}
\ha\ emission in Integral Field Spectroscopy data cubes of $\sim$1200 nearby galaxies obtained with PMAS and MUSE. We found 19 unique regions that pass our quality cuts, four of which match the locations of previously discovered SNe: one Type IIP, and three Type IIn, including the well-known SN 2005ip. We suggest that these objects are young Supernova Remnants, with bright and broad \ha\ emission powered by the interaction between the SN ejecta and dense circumstellar material. The stellar ages measured at the location of these SNR candidates are systematically lower by about 0.5 dex than those measured at the location of core collapse SNe, implying that their progenitors might be shorter lived and therefore more massive than a typical CC SN progenitor. The methods laid out in this work open a new window into the study of nearby SNe with Integral Field Spectroscopy. 
\end{abstract}


%%%%%%%%%%%%%%%%%%%%%%%%%%%%%%%%%%%%%%%%%%%%%%%%%%%%%%%%%%%%%%%%%%%%%
%%%%%%%%%%%%%%%%%%%%%%%%%%%%%%%%%%%%%%%%%%%%%%%%%%%%%%%%%%%%%%%%%%%%%
%%%%%%%%%%%%%%%%%%%%%%%%%%%%%%%%%%%%%%%%%%%%%%%%%%%%%%%%%%%%%%%%%%%%%

\section{Introduction} \label{sec:Introduction}

Supernovae (SNe) are energetic stellar explosions that mark the endpoints in the life of certain types of stars. Although they are rare events, occurring once or twice per century in a typical galaxy, SNe are essential to understanding the chemical evolution of the Universe \citep{Ko06, An16,Pr18} and the injection of energy into the interstellar medium \citep[ISM,][]{Tho98}. This local deposition of energy plays a crucial role in galaxy evolution, triggering star formation \citep{Sti06,DV12,Hop14} and seeding and sustaining turbulence \citep{MacLow2004}. 

Our understanding of the role that SNe play in the ecology of the ISM is limited by the many questions that remain open about their stellar progenitors. The baseline physical scenarios for the two major classes of SNe posit that core collapse (CC) SNe arise from gravitational collapse in the cores of stars more massive than $\sim8$ \msun, and Type Ia SNe are the aftermath of a thermonuclear runaway ignited in the central regions of massive CO white dwarfs (WDs) that are somehow destabilized by accretion in close binary systems. However, many important details of these scenarios remain obscure, including the role played by binary interactions and pre-explosion progenitor mass loss in CCSNe \citep{Smartt2009,Smith2014}, and the specific identity of the progenitors in SN Ia \citep{Maoz2014}. As a result, we have not been able to fully characterize fundamental properties that regulate the feedback mechanism into the ISM, like the distribution of progenitor masses and delay times \citep{Badenes2009,Jennings2014,Zapartas2017,Auchettl2019,Strolger2020,Castrillo2021} or the detailed nucleosynthetic yields \citep{Romano2010,Andrews2017}.

Recent advances in astronomical instrumentation have opened new opportunities to explore these issues. In particular, the wide availability of data from Integral Field Spectroscopy (IFS) makes it possible to study the properties of the host galaxies of SNe with an unprecedented level of detail, revealing a wealth of information on the environments in which SN explode (e.g. \citealt{Sta12,Kun13a,Kun13b,Rig13,Gal14a,Gal16a,Gal16b,Gal17,Kru17,Gal18a,Kun18,Lym18,Rig18,Lym20}). Previous IFS studies of SNe have relied on dedicated SN surveys and literature searches to identify SN host galaxies and locate their explosion sites. Here, we describe the first results of an effort to use the IFS observations themselves to discover recent SNe. This is possible because some SNe undergo strong interaction with a dense circumstellar medium (CSM) shortly after the explosion, leading to large luminosities that can be sustained for a long time \citep{Milisavljevic2017,Dessart2023}. These objects are often described as transitional, between SNe and Supernova Remnants (SNRs), and range from young SNe with clear evidence of CSM interaction such as SN 1993J \citep{Fransson1996,Matheson2000}, SN1996cr \citep{Bauer2008,Quirola-Vasquez2019}, and SN 1978K \citep{Kuncarayakti2016} to SNRs like Cas A, which remain optically bright centuries after the explosion \citep{Milisavljevic2012}. For clarity, we will refer to this class of objects as young SNRs. Our work is motivated by the fact that some of these young SNRs are luminous enough to stand out in narrow-band images of their host galaxies \citep[e.g., SNR NGC4449,][]{Milisavljevic2008}, and in principle could be picked up serendipitously in IFS observations of nearby galaxies. To explore this possibility, we have conducted a systematic search for regions in IFS data cubes that are characterized by bright \ha\ line emission with a significant broad component ($\geq$ 400 km s$^{-1}$) that could be associated with CSM interaction.

This paper is organized as follows. In Section \ref{sec:Observations}, we describe the IFS data that constitute our initial sample. In Section \ref{sec:Method}, we outline the methods we have employed to identify candidate young SNRs. In Section \ref{sec:Regions_interest}, we discuss the main properties of our sample of young SNR candidates. In Section \ref{sec:conclusions} we comment on our results and discuss possible avenues for future research.

\begin{figure*}[!t]
\centering
\includegraphics[width=\textwidth]{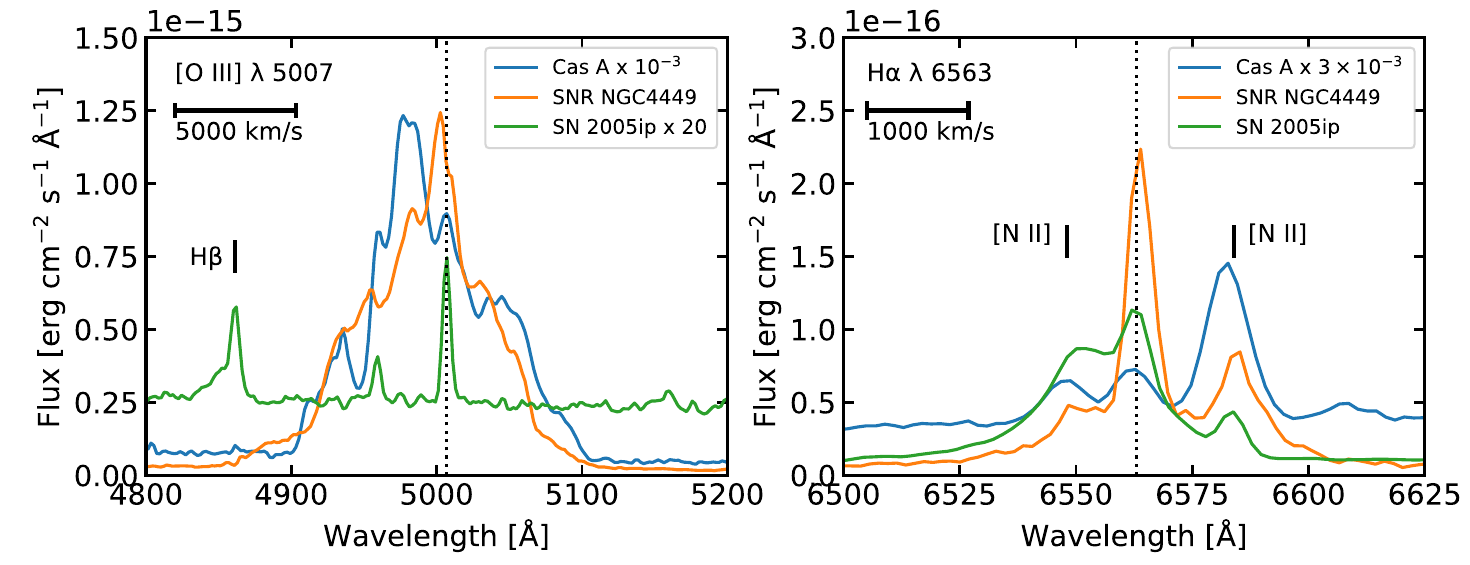}
\caption{Optical spectra of Cas A and SNR NGC4449 \citep[both from][]{Milisavljevic2012}, and SN 2005ip (from our PMAS IFS data) around the [O III] $\lambda$5007 line (left), and the \ha\ line (right).}
\label{fig:YoungSNRs}
\end{figure*}

%%%%%%%%%%%%%%%%%%%%%%%%%%%%%%%%%%%%%%%%%%%%%%%%%%%%%%%%%%%%%%%%%%%%%
%%%%%%%%%%%%%%%%%%%%%%%%%%%%%%%%%%%%%%%%%%%%%%%%%%%%%%%%%%%%%%%%%%%%%
%%%%%%%%%%%%%%%%%%%%%%%%%%%%%%%%%%%%%%%%%%%%%%%%%%%%%%%%%%%%%%%%%%%%%
\section{Observations} \label{sec:Observations}

Our initial sample is constituted by IFS observations of nearby galaxies obtained using two different instruments:
\begin{itemize}
\item[(i)] the Potsdam Multi Aperture Spectograph (PMAS; \citealt{2005PASP..117..620R}) in PPak mode \citep{2004AN....325..151V,2006PASP..118..129K}, mounted on the 3.5 m telescope of the Centro Astronomico Hispano-Aleman (CAHA) at the Calar Alto Observatory. PPak consists of a fiber bundle of 382 fibers with 2.7" diameter. Among these, 331 are ordered in a single hexagonal bundle, with the remaining fibers used for sky measurements and calibration purposes. Most objects are observed with two overlapping setups and then combined. The V500 grating has a spectral resolution of $\sim$6 \AA~in the wavelength range 3750$-$7300 \AA. The V1200 grating has a higher spectral resolution of $\sim$2.7 \AA, with a bluer range spanning 3400-4750 \AA. The final products are 3D datacubes with a 100\% covering factor within a hexagonal FoV of $\sim$1.3 arcmin$^2$ with 1"$\times$1" pixels, which correspond to $\sim$4000 spectra per object. 
\item[(ii)] the Multi-Unit Spectroscopic Explorer (MUSE; \citealt{2014Msngr.157...13B}), located at the Nasmyth B focus of Yepun, the VLT UT4 telescope at Cerro Paranal Observatory. It has a modular structure composed of 24 identical IFU modules that together sample, in Wide Field Mode (WFM), a near-contiguous 1 arcmin$^2$ FoV with spaxels of 0.2 $\times$ 0.2 arcsec, and a wavelength coverage of 4650-9300 \AA~with a mean resolution of R $\sim$3000. MUSE produces $\sim$100,000 spectra per pointing.
\end{itemize}

Observations obtained with PMAS come in their majority from the 3rd data release of the Calar Alto Legacy Integral Field Area survey (CALIFA; \citealt{San12,Sanc16}). The PMAS sample consists of 667 galaxies selected from DR7 of the Sloan Digital Sky Survey \citep[SDSS,][]{2009ApJS..182..543A} with redshifts between 0.005 and 0.03, declinations $>$-7$^\circ$, and diameters between 45" and 79.2". These selection criteria are optimized for the PMAS/PPak instrument. We added to this sample data from the PMAS/PPak Integral field Supernova hosts COmpilation (PISCO; \citealt{Gal18a}), an extended project of CALIFA that aimed at increasing the sample of supernova host galaxies used for environmental studies \citep{Gal14a,Gal16b}. As of January 2020, the PISCO sample contained 220 galaxies, bringing the total PMAS sample to 887 galaxies.

Observations obtained with MUSE come in its totality from the All-weather MUse Supernova Integral-field of Nearby Galaxies (AMUSING; \citealt{Gal16a}; Galbany et al. in prep.) survey. This survey has been running for 11 semesters (P95-P106) with the focus of obtaining IFS of SN host galaxies, each semester with a different science focus including among others: hosts of SNe that showed strong sodium absorption lines in their spectra indicating the presence of large amounts of dust; hosts of SNe discovered by the ASASSN-SN survey to study SN rates as a function of local environment \citep{2023A&A...677A..28P}; SN hosts with low surface brightness \citep{2022arXiv220707657H}; hosts of SNe included in the Carnegie Supernova Project (CSP; \citealt{2019PASP..131a4001P}) sample.
The compilation used in this work consisted of 342 nearby SN host galaxies. %\redpen{State specific number of galaxies and give a citation that describes selection criteria} 

%The selection of targets was driven by different science cases in each semester, e.g. SN Ia hosts, CC SNe hosts, etc.

%%%%%%%%%%%%%%%%%%%%%%%%%%%%%%%%%%%%%%%%%%%%%%%%%%%%%%%%%%%%%%%%%%%%%
%%%%%%%%%%%%%%%%%%%%%%%%%%%%%%%%%%%%%%%%%%%%%%%%%%%%%%%%%%%%%%%%%%%%%
%%%%%%%%%%%%%%%%%%%%%%%%%%%%%%%%%%%%%%%%%%%%%%%%%%%%%%%%%%%%%%%%%%%%%

\section{Detection method} \label{sec:Method}

\subsection{Pre-processing} \label{subsec:Pre-processing}

All IFS datacubes are pre-processed in preparation for the analysis. First, regions with foreground stars are masked out. Then, single stellar population (SSP) synthesis models are fit to all individual spectra (around 4,000 and 100,000 spectra per PMAS and MUSE cube, respectively) to separate the underlying stellar continuum from the ionized gas-phase emission. This is done with {\tt STARLIGHT} \citep{2005MNRAS.358..363C,2009RMxAC..35..127C}, which determines the fractional contribution of different SSP models to the spectrum, $x_i$, and to the galaxy mass, $\mu_i$. {\tt STARLIGHT} accounts for dust extinction ($A^*_V$) as a foreground screen, using a reddening law from \cite{1989ApJ...345..245C} that assumes $R_V$=3.1. To reduce computing time, we used a reduced grid of 248 SSP models from \cite{2013A&A...557A..86C}, with 62 ages spanning 1~Myr to 14~Gyr and four metallicities (0.2, 0.4, 1.0, and 1.5 solar). The best fit model is subtracted from each observed spectrum and the resulting residual is stored in a 3D cube containing only the ionized gas-phase emission.

\begin{figure*}[!t]
\centering
\includegraphics[trim=0.75cm 0.75cm 0.75cm 0.75cm, clip=True,width=0.49\textwidth]{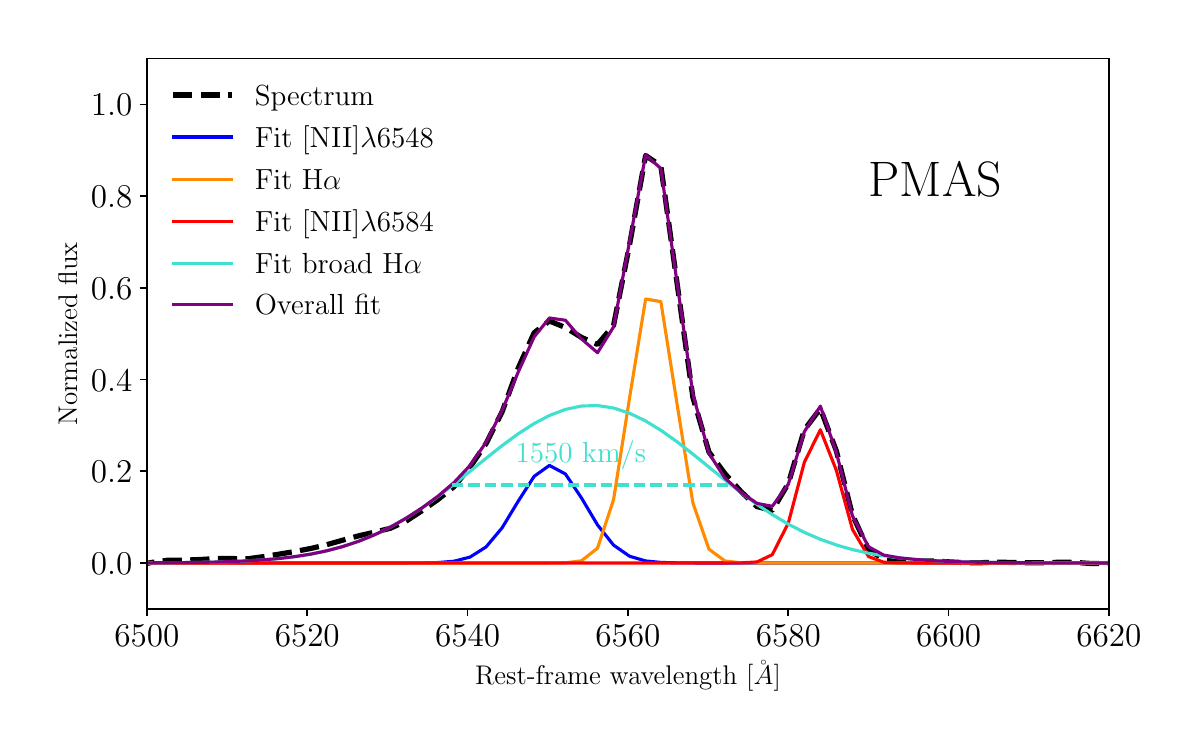}
\includegraphics[trim=0.75cm 0.75cm 0.75cm 0.75cm, clip=True,width=0.49\textwidth]{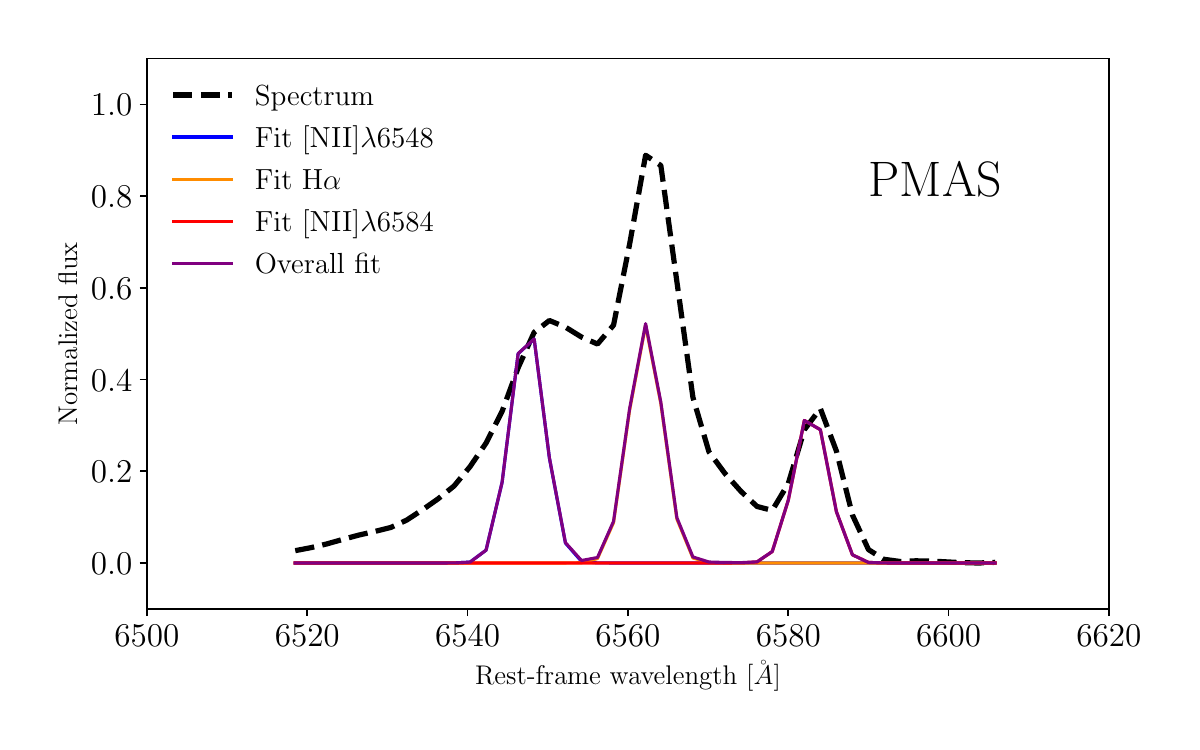}
\includegraphics[trim=0.75cm 0.75cm 0.75cm 0.75cm, clip=True,width=0.49\textwidth]{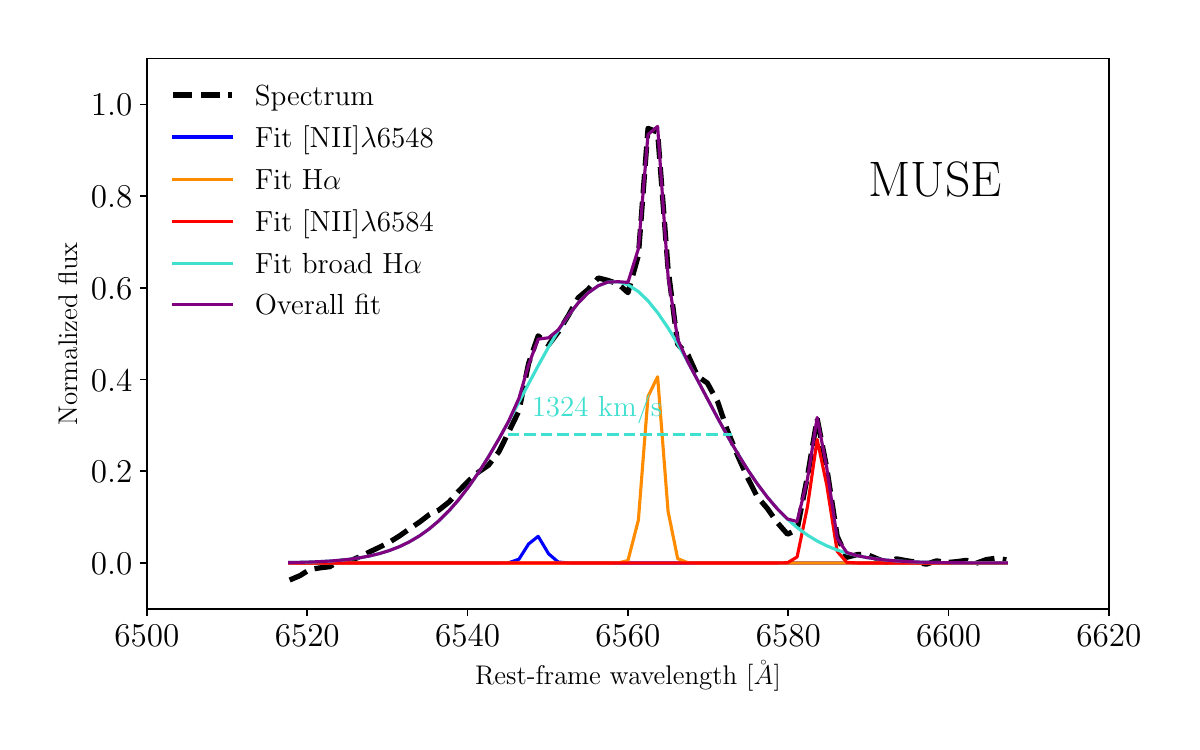}
\includegraphics[trim=0.75cm 0.75cm 0.75cm 0.75cm, clip=True,width=0.49\textwidth]{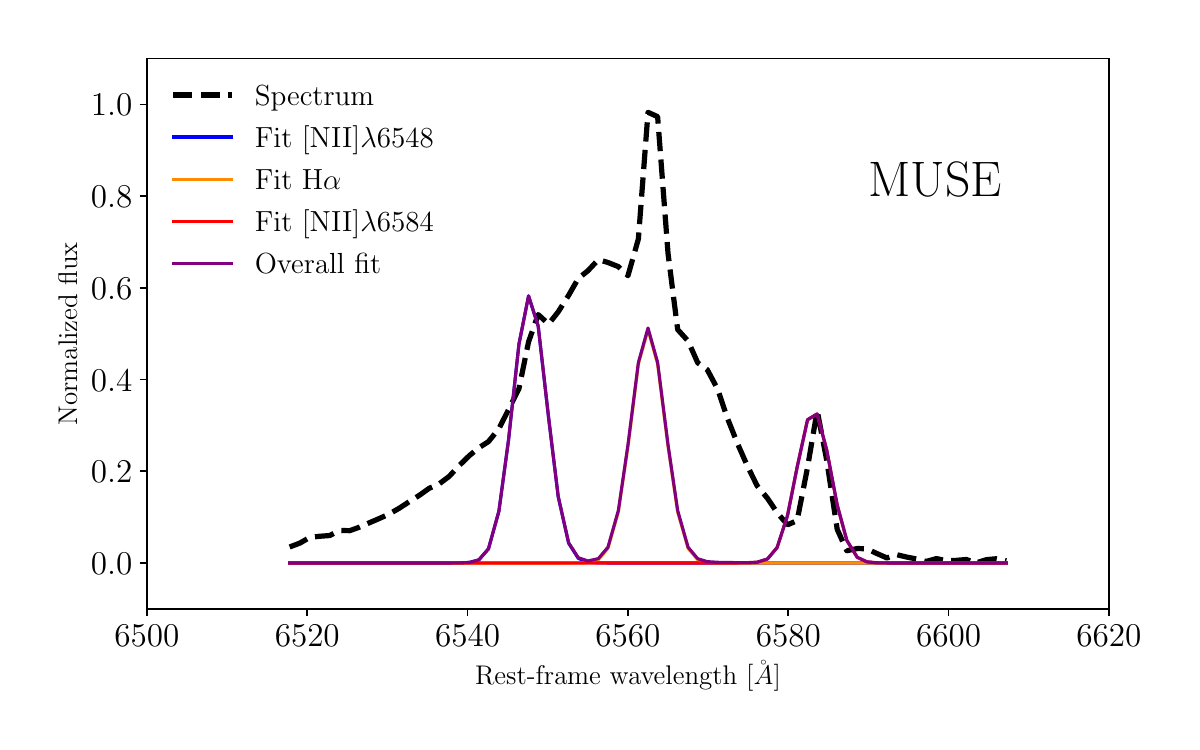}
\caption{\textit{Left:} Line fits to the residual spectrum of NGC 2906 at the location of SN~2005ip in PMAS (top) and MUSE (bottom). Blue, red and orange lines show the narrow [\ion{N}{ii}] doublet and \ha, while turquoise shows the broad \ha\ component corresponding to SN 2005ip. This broad component is blueshifted with respect to the narrow \ha\ emission from the host galaxy. The full width at half maximum (FWHM) of the \ha\ emission from SN 2005ip is shown as the horizontal dashed line. The overall fit and emission spectrum are shown with purple and a black, dashed lines. \textit{Right}: The same residual spectra shown in the left panels, fitted without including the broad \ha\ component. Notice the inability of the fitted narrow line emission (purple) to reproduce the observed residual spectrum (dashed black).}
\label{fig:4thHalpha}
\end{figure*}

%\redpen{Did you fit all the lines in the residual spectra, or only the ones you used to look for SNRs? If that is the case, this paragraph belongs in the next section. In any case, you should explicitly say how many lines you fitted, list their wavelengths and explain why you chose these, and not others}.
The resulting residual spectra are mostly flat, dominated by the narrow line emission from the host galaxy. We fit the most prominent lines (H$\beta$, [\ion{O}{iii}] $\lambda$5007, [\ion{O}{iii}] $\lambda$4969, [\ion{N}{ii}] $\lambda\lambda$6548,84, H$\alpha$, [\ion{S}{ii}] $\lambda\lambda$6719,31) with the \texttt{Python} version of \mpfit\ \citep{Mar09,Mar12,Ne16}, which performs non-linear least-squares fitting using the Levenberg-Marquardt minimization algorithm \citep{Lev44,Mar63}. For each residual spectrum, we focus on the wavelength range corresponding to each line and fit it using a Gaussian profile to retrieve line fluxes, central wavelengths, velocity shifts and widths, 
%\redpen{What, exactly, do you fit for? Isn't width and velocity the same thing?}
and their corresponding errors. When required, more than one Gaussian is fitted simultaneously for transitions that are close to one another (H$\alpha$ with the two [\ion{N}{ii}], the two [\ion{S}{ii}], and H$\beta$ with [\ion{O}{iii}]).
%\redpen{This paragraph is very unclear. I have no idea what you did}

We correct all the fitted line fluxes for reddening intrinsic to the host galaxy by measuring the \ha\ to \hb\ ratio and applying an extinction law from \citet{1989ApJ...345..245C}, assuming $R_{V} = 3.1$, Case B recombination, and densities of $\sim$10$^3$ cm$^{-3}$ around a heating source with T$\sim$10$^4$ K and a large optical depth \citep{2006agna.book.....O}. We note that the extinction correction and the selection of a particular extinction law do not affect the method we use to detect young SNR candidates.

\subsection{Line emission from young SNRs in IFS data: an illustrative example} \label{subsec:broad_Ha}

Young SNRs have distinctive optical spectra. In Figure \ref{fig:YoungSNRs} we show the integrated line emission in two spectral windows around the [\ion{O}{iii}] $\lambda$5007 and \ha\ lines from the Galactic SNR Cas A (age $\sim$340 yr, \citealt{Thorstensen2001}, \citealt{Milisavljevic2012}, the SNR in NGC 4449 (age $\sim$80 yr, \citealt{Milisavljevic2008}, \citealt{Bietenholz2010}), and SN~2005ip (age $\sim$17 yr), which we use as an illustrative example of a young SNR imaged by our IFS data. SN~2005ip is a bright, well-observed Type IIn SN that exploded in NGC 2906 \citep{Fox09,Fox10} and has been showing signs of strong CSM interaction and enhanced dust formation for well over a decade \citep{Sm09,Str12,Kat14,Sm17,Bevan2019,Fox2020}. 
%The line emission from these three objects has a few salient features that can be exploited to conduct a systematic search for young SNRs in IFS data. 
All three objects show bright emission in both lines, with Cas A and SNR NGC4449 being brighter in [\ion{O}{iii}], and SN~2005ip being brighter in \ha. The \ha\ emission is noticeably broadened by several hundred \kms\ due to shock interactions in all three objects, although in all cases this broad emission appears superimposed on several components of narrow emission from \ha\ and the neighboring [\ion{N}{ii}] $\lambda\lambda$6548,84 doublet. The [\ion{O}{iii}] $\lambda$5007 line is broadened by several thousand \kms\ in both Cas A and SNR NGC4449, but the spectrum from SN~2005ip appears much narrower, even if the neighboring \hb\ line has a clear blueshifted broad component.

%\redpen{Danny Milisavljevic might be able to provide better example spectra.}

%\textcolor{green}{I think this is good, becasue the point here is it comes from IFS}.

\begin{figure}
\centering
\includegraphics[width=\columnwidth]{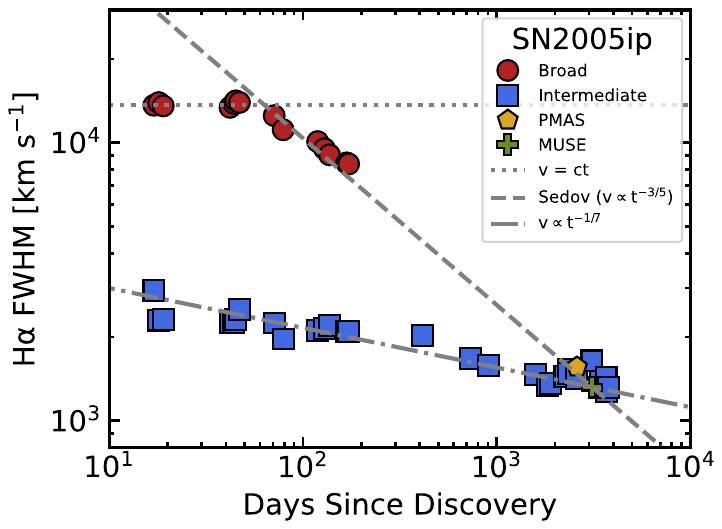}
\caption{Temporal evolution of the FWHM for the two components of the \ha\ emission from SN~2005ip, calculated from the spectra analyzed in \cite{Sm09} and \cite{Sm17} (red circles for the broad component, blue squares for the intermediate component), together with the IFS observations from PMAS and MUSE shown in Figure~\ref{fig:4thHalpha} (yellow pentagon and green cross). Three lines are included for illustrative purposes: a constant velocity set to the FWHM of the broad component in the first spectrum ($t=17$ days), a Sedov solution ($v \propto t^{-3/5}$) anchored to the FWHM of the MUSE spectrum, and a $v \propto t^{-1/7}$ powerlaw, also anchored to the FWHM of the MUSE spectrum.}
\label{fig:SN2005ip_Ha_FWHM}
\end{figure}

A detailed physical interpretation of the line emission from young SNRs is outside the scope of the present work. In particular, the mapping between shock physics and line emission is likely to be complex - we note that our IFS data only provide an unresolved view of objects that probably have a great deal of spatial structure, which can only be studied in detail in nearby cases like Cas A \citep{Milisavljevic2012,Milisavljevic2015}. We refer the reader to \cite{Fransson2002} and  \cite{Milisavljevic2012} for further details. For our purposes, the qualitative comparisons shown in Figure \ref{fig:YoungSNRs} are enough to illustrate the ability of IFS data cubes to recover young SNe with strong CSM interaction.

\begin{table}[!t]
\centering
\caption{Full Width Hal Maximum of the two components, broad and intermediate, of SN~2005ip measured in spectra obtained from the literature plus our PMAS and MUSE measurement.\label{tab:sn2005ip}}
\begin{tabular}{lcc}
\hline \hline
Date & FWHM$_{\rm Broad}$ & FWHM$_{\rm Intermediate}$ \\
MJD & [km s$^{-1}$] & [km s$^{-1}$] \\
\hline
53697 & 13625.0 $\pm$ 1899.9 & 2936.1 $\pm$ 1982.2 \\
53698 & 13936.1 $\pm$ 2031.7 & 2299.7 $\pm$ 1302.9 \\
53699 & 13556.2 $\pm$ 1865.8 & 2310.0 $\pm$ 1218.9 \\
53722 & 13393.3 $\pm$ 1891.5 & 2250.6 $\pm$ 1360.3 \\
53724 & 13841.3 $\pm$ 2005.5 & 2245.7 $\pm$ 1399.2 \\
53725 & 14133.7 $\pm$ 1994.1 & 2317.4 $\pm$ 1472.2 \\
53727 & 13947.9 $\pm$ 1910.5 & 2508.1 $\pm$ 1663.6 \\
53751 & 12507.7 $\pm$ 1544.1 & 2245.2 $\pm$ 1128.8 \\
53759 & 11121.5 $\pm$ 1325.6 & 1965.7  $\pm$ 714.0 \\
53799 & 10085.8 $\pm$ 1587.2 & 2113.0  $\pm$ 848.7 \\
53809 &  9507.8 $\pm$ 538.9 & 2142.5  $\pm$ 525.7 \\
53817 &  9037.8  $\pm$ 670.6 & 2204.2  $\pm$ 670.6 \\
53848 &  8458.0 $\pm$ 681.1 & 2099.8  $\pm$ 681.1 \\
53852 &  8272.0 $\pm$ 703.5 & 2093.0  $\pm$ 488.3 \\
54092 &     $...$        & 2026.7  $\pm$ 339.4 \\
54415 &     $...$        & 1677.9  $\pm$ 364.9 \\
54584 &     $...$        & 1579.7  $\pm$ 302.2 \\
55267 &     $...$        & 1466.1  $\pm$ 331.9 \\
55510 &     $...$        & 1332.2  $\pm$ 342.5 \\
55574 &     $...$        & 1361.9  $\pm$ 264.4 \\
55928 &     $...$        & 1444.0  $\pm$ 328.0 \\
56030 &     $...$        & 1522.7  $\pm$ 379.8 \\
56033 &     $...$        & 1477.5  $\pm$ 351.7 \\
56255 &     $...$        & 1421.3  $\pm$ 262.1 \\
56273 &     $...$        & 1559.6  $\pm$ 636.9 \\
56778 &     $...$        & 1639.7  $\pm$ 392.9 \\
56783 &     $...$        & 1320.4  $\pm$ 350.6 \\
57103 &     $...$        & 1381.5  $\pm$ 321.0 \\
57346 &     $...$        & 1430.7  $\pm$ 352.9 \\
57372 &     $...$        & 1273.6  $\pm$ 238.6 \\
57433 &     $...$        & 1314.2  $\pm$ 262.8 \\
57449 &     $...$        & 1318.3  $\pm$ 310.4 \\
\hline
\end{tabular}
\end{table}

Detailed fits to the line emission from the PMAS and MUSE residual spectra around \ha\ at the location of SN~2005ip are shown in Figure \ref{fig:4thHalpha}. The default fits to the narrow line emission from the host galaxy are shown on the right panels. The three narrow Gaussians for \ha\ and the [\ion{N}{ii}] doublet emission miss much of the measured flux, and try to compensate for this by overfitting the [\ion{N}{ii}] line at $\lambda=$ 6548 \AA, resulting in an unphysical flux ratio for the [\ion{N}{ii}] doublet. The addition of a fourth Gaussian corresponding to a broad \ha\ component associated with SN~2005ip greatly improves the spectral fit, accounting for all the observed flux and recovering the correct flux ratio for the narrow component of the [\ion{N}{ii}] doublet. The width of this component in the IFS data cubes is $\sigma_{\rm PMAS} = 14.41 \pm 5.92 \,$\AA~and $\sigma_{\rm MUSE} = 12.31 \pm 3.26 \,$\AA, which correspond to FWHM velocities of 1550$\pm$637 and 1324$\pm$351 \kms, respectively. 

To evaluate our IFS observations of SN~2005ip in context, we have compiled the 30 single-slit spectra of the SN analyzed by \cite{Sm09} and \cite{Sm17}, which span ages between 17 days and more than 10 years after discovery. As noted in these references, the broad \ha\ emission from SN~2005ip has at least two components, which evolve in distinct ways (see e.g. Figure 5 in \citealt{Sm09}). In Table~\ref{tab:sn2005ip}, we list fitted FWHM values for these two components, which we have labeled 'broad' and 'intermediate', in the single-slit spectra of SN~2005ip. The overall temporal evolution of these FWHM values is shown in Figure~\ref{fig:SN2005ip_Ha_FWHM}, together with our IFS observations. The broadest of the two components is only seen clearly in spectra taken before April 2006 (MJD 53852, or 172 days after discovery, see Table~\ref{tab:sn2005ip}), but the intermediate component is always present. The evolution of the broad component shows an initial free expansion ($v \sim$ ct) phase, with a clear transition to a Sedov regime ($v \propto t^{-3/5}$) approximately 60 days after the explosion. This early transition to a Sedov regime, which happened while the SN was still optically bright, is consistent with the hypothesis that the SN progenitor lost a large amount of mass shortly before the explosion \citep[see][for a discussion]{Moriya2013}. The intermediate component, on the other hand, shows a slower deceleration, roughly as $v \propto t^{-1/7}$. The physical interpretation for the behavior of the intermediate component is less clear, but we emphasize that our IFS measurements are broadly consistent with those obtained from contemporary single-slit spectra, showcasing the ability of galaxy-wide IFS data sets to study bright young SNRs and obtain physically meaningful measurements from them.

\begin{figure}[!t]
\centering
\includegraphics[width=\columnwidth]{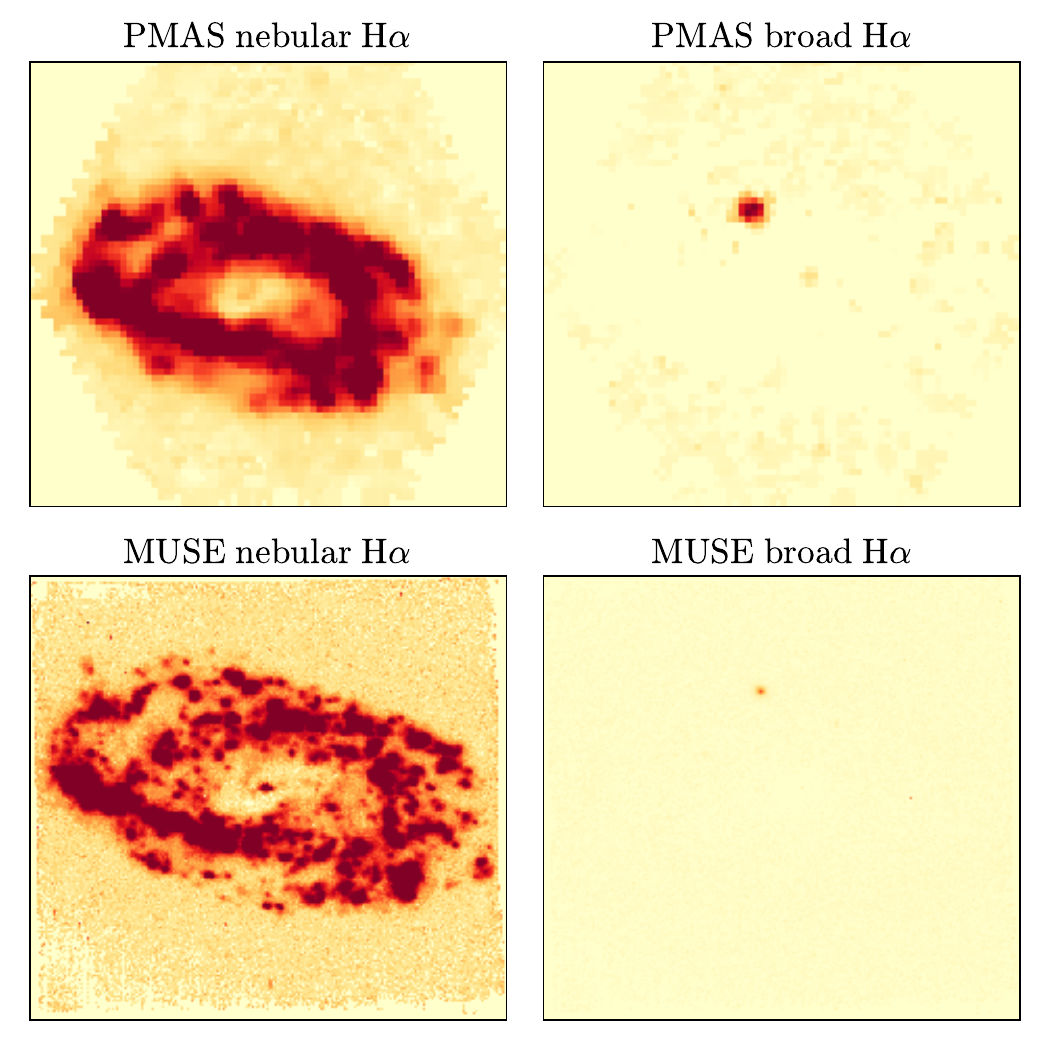}
\caption{H$\alpha$ flux maps for NGC~2906. The \ha\ contribution from SN 2005ip (\textit{right}) can be separated from that of its host galaxy (\textit{left}). \textit{Top}: PMAS. \textit{Bottom}: MUSE.}
\label{fig:4thHa_maps}
\end{figure}

%and confirm that the width has remained roughly stable since the lightcurve reached a plateau $\sim200$ days after the explosion. \redpen{We might want to add the MUSE and PMAS measurements to the data points shown in Figure 15 of \cite{Str12} to emphasize that we can get physical information from these observations. I can try to make that plot if you send me the date at which the data was taken.}

\begin{figure*}[!t]
\centering
\includegraphics[width=0.69\textwidth]{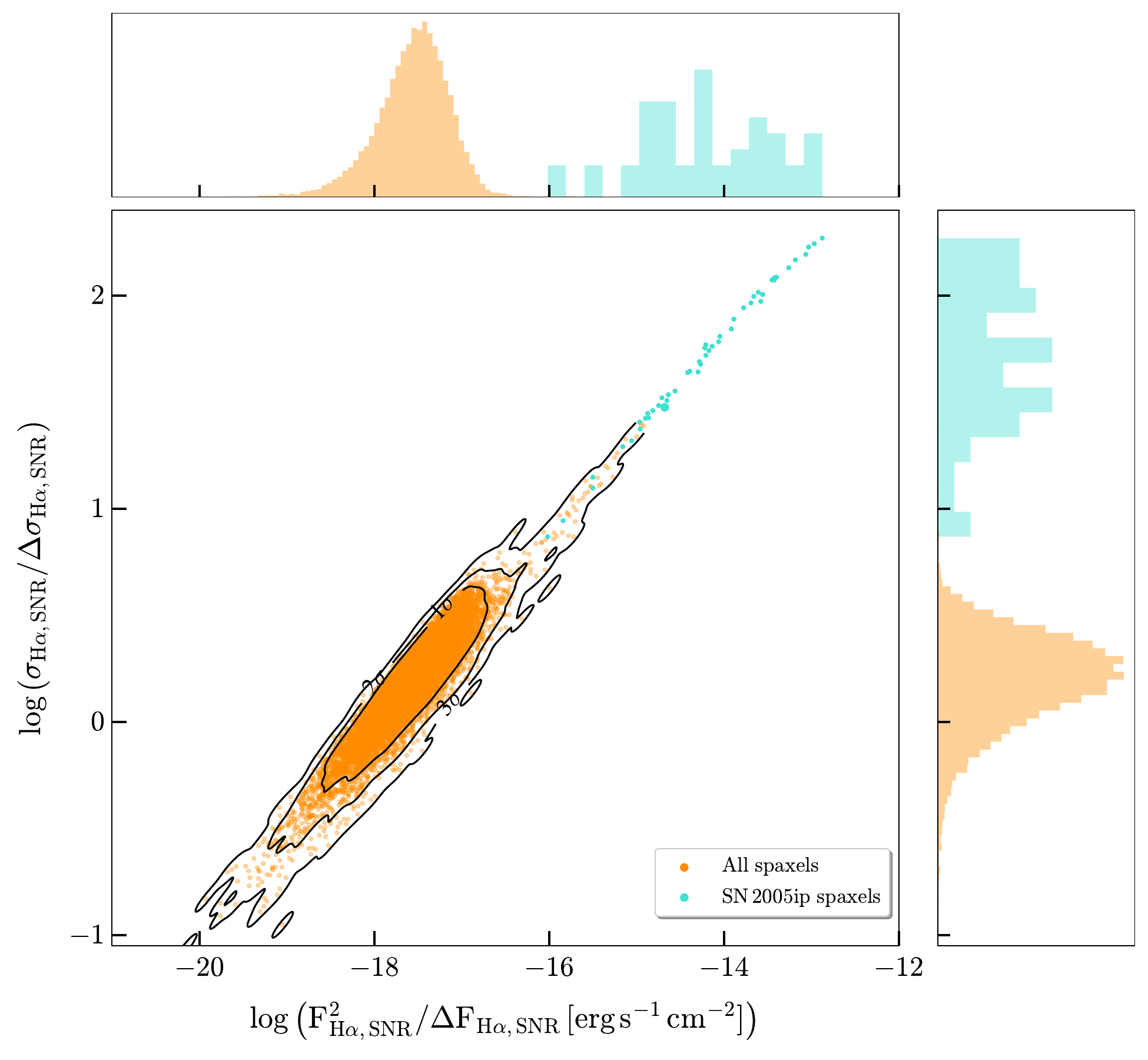}
\caption{Diagnostic plot showing the MUSE data for the entire galaxy NGC2906. $F$, $\Delta F$, $\sigma$, and $\Delta \sigma$ represent the fluxes and widths of the broad components in the \ha\ line with their uncertainties. Spaxels where the broad H$\alpha$ component was detected are shown in turquoise, while the rest of the spaxels in the galaxy are shown in orange. The black contours indicate the 1-, 2- and 3-$\sigma$ confidence contours of the distribution of each quantity.}
\label{fig:impactfactor}
\end{figure*}

\begin{table*}
\footnotesize
\begin{center}
\caption{Coordinates, broad H$\alpha$ fluxes and luminosities of the 20 SNR candidate regions with broad H$\alpha$ emission.}
\label{table:group1_broadHa_fluxes}
\begin{tabular}{lccccccc}
Host Galaxy & Redshift & Survey & RA & DEC & $\rm{v_{H\alpha, SNR}}$ & $\rm{FWHM_{H\alpha, SNR}}$ & $\rm{L_{H\alpha, SNR}}$ \\
 & & & $\mathrm{J2000}$ &  $\mathrm{J2000}$& [$\mathrm{km/s}$] & [$\mathrm{km/s}$] & [$\mathrm{10^{36} \, ergs/s}$] \\ 
\hline
%\multicolumn{8}{c}{SNR H$\alpha$} \\ 
\hline
LEDA 43070\tablenotemark{a} & 0.0158 &  MUSE  & 12:46:14.75  & -40:48:52.59 & $141\pm63$   & $1932\pm547$ & $3.159\pm0.760$\\
2MASX J23331223-6034201 & 0.0148 &  MUSE  & 23:33:10.63  & -60:34:29.87 & $103\pm86$ & $442\pm216$  & $1.206\pm0.212$\\
LEDA 1015413 & 0.0131 &  MUSE  & 12:54:12.58  & -07:38:37.19 & $161\pm62$   & $503\pm358$  & $10.811\pm2.018$\\
NGC 2906\tablenotemark{b} & 0.0074 &  MUSE  & 09:32:06.42  & +08:26:44.53 & $-258\pm53$  & $1325\pm350$ & $7.720\pm0.806$\\
NGC 4806\tablenotemark{c} & 0.0082 &  MUSE  & 12:56:14.02  & -29:29:54.74 & $-11\pm51$ & $953\pm358$  & $7.551\pm1.891$ \\ % SN2011fh   
ESO 467-51\tablenotemark{d} & 0.0060 &  MUSE  & 22:23:16.15  & -28:58:31.17 & $288\pm90$ & $1551\pm299$ & $7.045\pm1.023$ \\ % ASASSN-14jb
NGC 1448 & 0.0037 &  MUSE  & 03:44:30.39  & -44:38:50.48 & $157\pm25$   & $1504\pm485$ & $0.290\pm0.046$ \\ % SN2001el   
Arp 142 & 0.0233 &  PMAS  & 09:37:42.50  & +02:45:26.76 & $-226\pm73$  & $415\pm209$  & $29.628\pm7.400$ \\
MCG +11-08-25 & 0.0136 &  PMAS  & 06:15:48.81  & +66:50:21.60 & $174\pm68$   & $1271\pm386$ & $5.933\pm1.744$\\
NGC 1056 & 0.0052 &  PMAS  & 02:42:47.98  & +28:34:27.41 & $286\pm23$   & $1661\pm489$ & $0.398\pm0.055$\\
NGC 2276 & 0.0081 &  PMAS  & 07:27:07.12  & +85:44:35.71 & $-411\pm78$  & $411\pm261$  & $1.327\pm0.264$ \\
NGC 2906\tablenotemark{b} & 0.0074 &  PMAS  & 09:32:06.47  & +08:26:44.94 & $-334\pm42$  &  {\bf $1552\pm638$} & $4.2513\pm0.510$ \\ 
NGC 5633 & 0.0078 &  PMAS  & 14:27:27.69  & +46:08:36.14 & $-201\pm84$ & $414\pm206$  & $1.579\pm0.179$ \\
NGC 5735 & 0.0125 &  PMAS  & 14:42:34.95  & +28:43:38.18 & $-284\pm78$ & $439\pm199$  & $4.997\pm0.590$ \\
NGC 5908\tablenotemark{e} & 0.0110 &  PMAS  & 15:16:42.08  & +55:24:59.85 & $679\pm79$   & $3423\pm713$ & $1.020\pm0.156$ \\
NGC 6946 & 0.0001 &  PMAS  & 20:34:51.87  & +60:12:52.70 & $-163\pm76$  & $480\pm276$  & $0.001\pm0.001$ \\ % NGC6946P3
UGC 04179 & 0.0186 &  PMAS  & 08:02:07.59  & +00:48:27.79 & $-581\pm61$  & $1486\pm551$ & $20.501\pm7.007$ \\
UGC 09110 & 0.0156 &  PMAS  & 14:14:12.55  & +15:37:04.78 & $501\pm20$   & $1311\pm421$ & $2.372\pm0.255$ \\
UGC 09182 & 0.0155 &  PMAS  & 14:20:45.95  & +21:56:16.12 & $-199\pm34$  & $401\pm121$  & $14.092\pm8.745$ \\
MCG +03-31-094 & 0.0030 &  PMAS &  12:17:26.07  & +17:39:16.06 & $-59\pm91$   & $1360\pm343$ & $0.109\pm0.012$\\
\hline
%\vspace{-0.8cm}
\end{tabular}
\end{center}
\vspace{-0.3cm}
\tablenotetext{a}{ASAS-SN 14fd \citep{Holoein2019}}
\tablenotetext{b}{SN 2005ip \citep{Fox09,Sm09}}
\tablenotetext{c}{SN2011fh \citep{Pessi2022}}
\tablenotetext{d}{ASAS-SN 14jb \citep{Holoein2019}}
\tablenotetext{e}{Coincides with SN impostor reported in 2012, PSNJ15164204+5525011 \citep{Benetti2012}}
\end{table*}

%\textcolor{green}{MUSE: 2014-05-06 23:32:13.411, 56783.98071078 PMAS: 2012-12-12 03:13:43, 56273.13511741}. 

%This component has been observed in other objects, and is thought to arise from either shocked gas or unshocked SN ejecta photoionized by shock radiation \citep{Fransson2002}.

%\redpen{Maybe a couple of sentences about broad line emission in IFS data cubes - where is it found? is it usually at the center? what lines are broadened? is it always AGN? references? yes, at the center, usually Ha+NII, but also LINERS, 'LIERS' (not nuclear), and Seyfert emission..

The broad component to the \ha\ emission associated with SN~2005ip does not appear at any other location in the IFS data cubes for the host galaxy NGC 2906. To illustrate this, we added a broad emission component to the fits to the residual spectra in the \ha\ region for all spaxels, in addition to the narrow \ha\ and the two narrow [\ion{N}{ii}] lines. This broad component has a minimum width of 400 \kms, but no minimum amplitude, so that MPFIT returns a zero value for the amplitude when the spectral fit does not require it. {\bf For the reminder of the paper, we will not distinguish between 'broad' and 'intermediate' components to the \ha\ line emission, as defined in the case of SN~2005ip, but will use the notation 'broad' to describe any contribution to the line flux that is clearly broader than usual (i.e., that is not kinematically narrow).} Figure \ref{fig:4thHa_maps} shows the flux maps for NGC 2906 generated by this procedure for both the narrow (left column) and broad (right column) \ha\ components, in the PMAS (top row) and MUSE (bottom row) data. The narrow component is distributed throughout the entire disk of the galaxy, showcasing emission from individual HII regions, but the broad component is restricted exclusively to the location of SN~2005ip. Note how the higher spatial resolution of MUSE (0.2" compared to 1" for PMAS) allows to accurately pinpoint the site of the SN. 

Without spatial coincidence with a previously recorded SN, as seen here for SN~2005ip, it is impossible to determine with absolute confidence that a region showing broad line emission in an IFS data cube is in fact a young SNR. In general, broad line emission in IFS data cubes is often associated with Active Galactic Nuclei in the central regions  (see e.g. \citealt{2013A&A...555L...1P,2013A&A...558A..43S,2020MNRAS.492.3073L}). Off-center broad line emission has been associated with sources like stellar outflows from  supergiants or Wolf-Rayet clusters (e.g. \citealt{Diaz1987}, \citealt{Terlevich1991}, \citealt{Kehrig2020}), but the vast majority of these examples have low luminosities, which would be hard to disentangle from the background in IFS data cubes, and modest widths, with FWHM below 200 \kms. In rare cases, bright \ha\ emission with FWHM in excess of 1000 \kms\ has been reported in single-slit spectra, like the giant HII region NGC 5471 in M101 \citep{Castaneda1990}. For this specific example, follow-up X-ray observations indicate the presence of at least one SNR at this location \citep{Williams1995}. Of course, larger scale ionized flows outflows have been found in starburst and post-starburst galaxies like the `green peas' \citep{RodriguezDelPino2019,Amorin2012,Hogarth2020}, but these outflows will not appear as high contrast point sources like the IFS detection of SN~2005ip shown in Figure~\ref{fig:4thHa_maps}.

\section{A systematic search for young SNRs in IFS data} \label{sec:Regions_interest}

\subsection{Search method}

%\textcolor{red}{The 400 km/s criterion is not mentioned anywhere in the method. Need to explain also why that number is chosen.}

The illustrative example of SN 2005ip demonstrates that IFS data cubes have the potential to both recover emission from young SNRs and extract relevant physical information about them. Motivated by this realization, we have conducted a systematic search for broad \ha\ emission in all the IFS data of all the nearby galaxies in our sample. To do this, we defined a set of criteria designed to single out young SNR candidates without making strong assumptions about the specific properties of their line emission. 
%We focus on the [\ion{O}{iii}] $\lambda$5007 and \ha\ lines, and fit the residual spectra for all the spaxels in our IFS data with a model that adds a broad \ha\ component to the set of narrow emission lines from the ionized gas phase in the host galaxies described above. 
First, we require a minimum FWHM of $400$ \kms, which should be enough to remove most non-SN local outflows.
Second, we use the flux ($F$), flux uncertainty ($\Delta F$), width ($\sigma$), and width uncertainty ($\Delta\sigma$) of the Gaussian fits to the broad \ha\ component (denoted as \ha,SNR) to define two diagnostic ratios designed to single out bright line emission from young SNR candidates:
\begin{itemize}
\item $\log \left(\sigma_{\rm H\alpha, SNR} / \Delta \sigma_{\rm H\alpha, SNR}\right)$ 
%\item $\sigma_{\rm [O III]}$ vs $\log \left(\sigma_{\rm [O III]} / \Delta \sigma_{\rm [O III]}\right)$ 
%\item $\log \left(\sigma_{\rm H\alpha, SNR}^{ 2} \times \Delta \sigma_{\rm H\alpha, SNR} \right)$ vs $\log \left(\sigma_{\rm [O III]] \,}^{\, 2} \times \Delta \sigma_{\rm [O III]} \right)$ 
\item $\log \left(F_{\rm H\alpha, SNR}^{ 2} / \Delta F_{\rm H\alpha, SNR} \right)$
%\item $\log \left(F_{\rm [O III] \,}^{\, 2} / \Delta F_{\rm [O III]} \right)$. 
\end{itemize}
%We found that the presence of a broad [\ion{O}{iii}] $\lambda$5007 component was not a good diagnostic  

%\input{Table_final.tex}

\begin{figure*}[!t]
\centering
\includegraphics[height=\textheight]{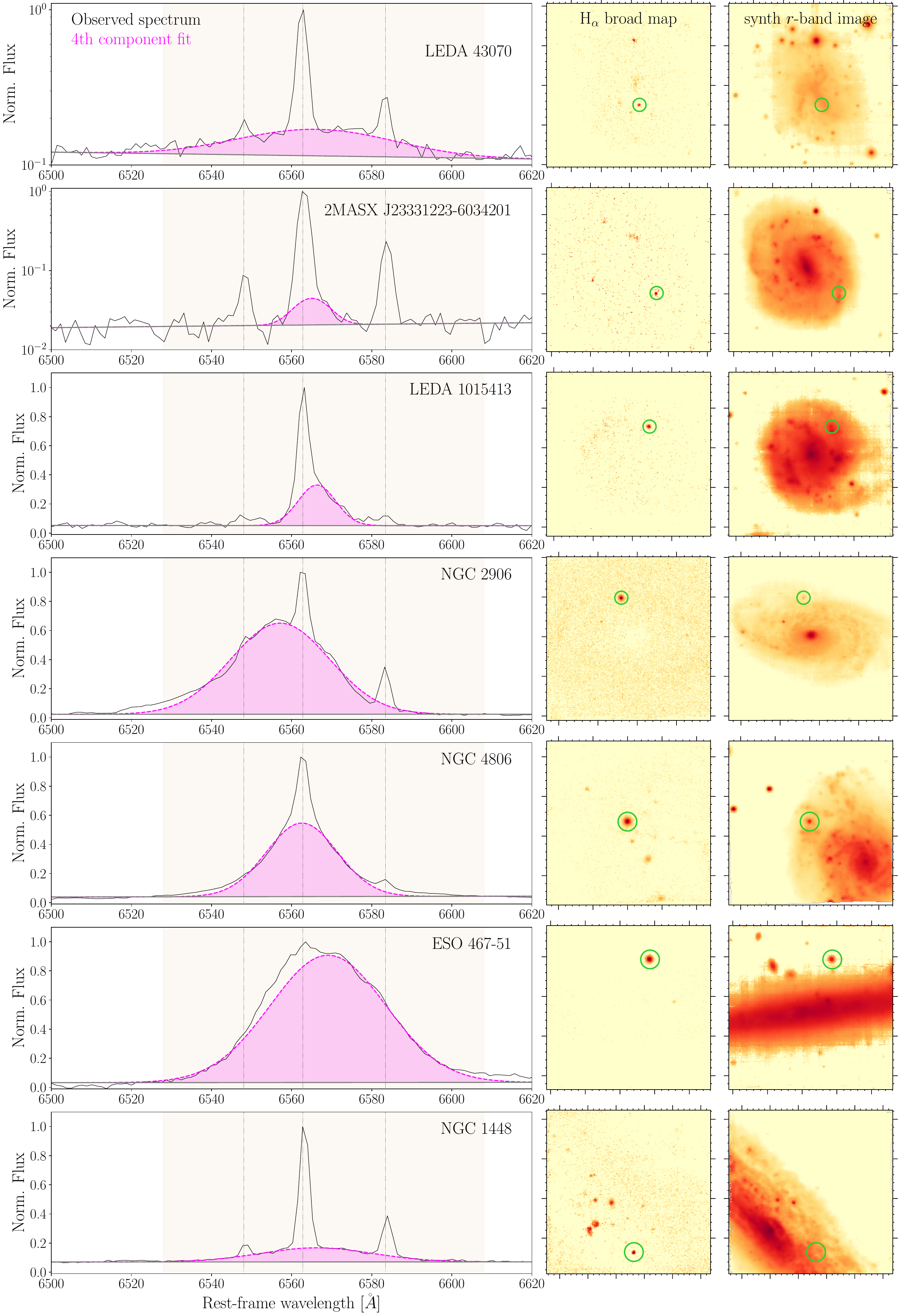}
\caption{Left column: spectra of young SNRs identified in our IFS sample in the \ha\ region, with the broad \ha\ component highlighted in magenta. Middle column: spatial distribution of the broad \ha\ component in the entire footprint of the IFS data. Right column: image of the IFS data footprint in white light.}
\label{fig:group1_all_1}
\end{figure*}

\begin{figure*}[!t]
\centering
\includegraphics[height=\textheight]{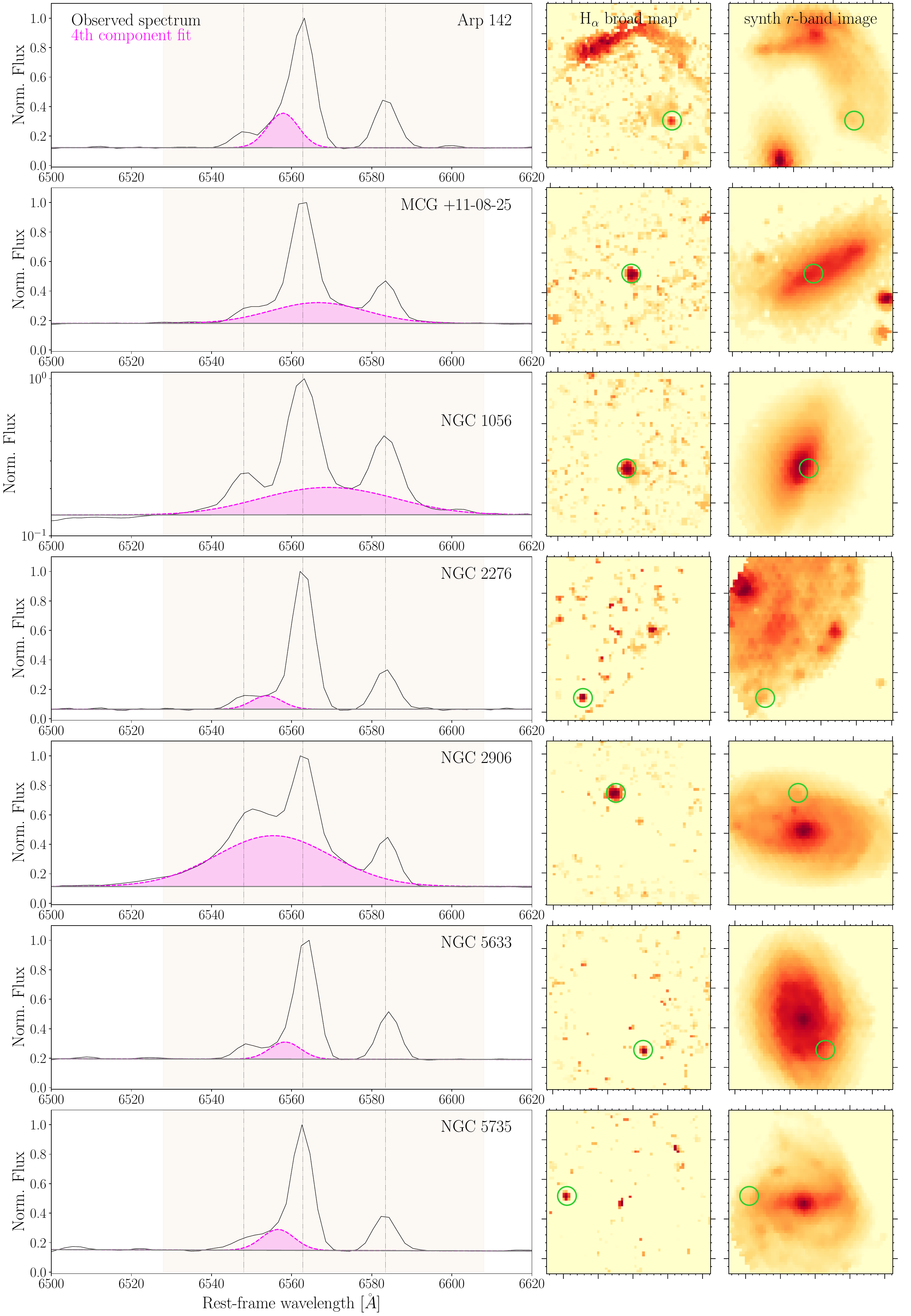}
\caption{(Same as Figure~\ref{fig:group1_all_1})}
\label{fig:group1_all_2}
\end{figure*}

\begin{figure*}[!t]
\centering
\includegraphics[width=0.9\textwidth]{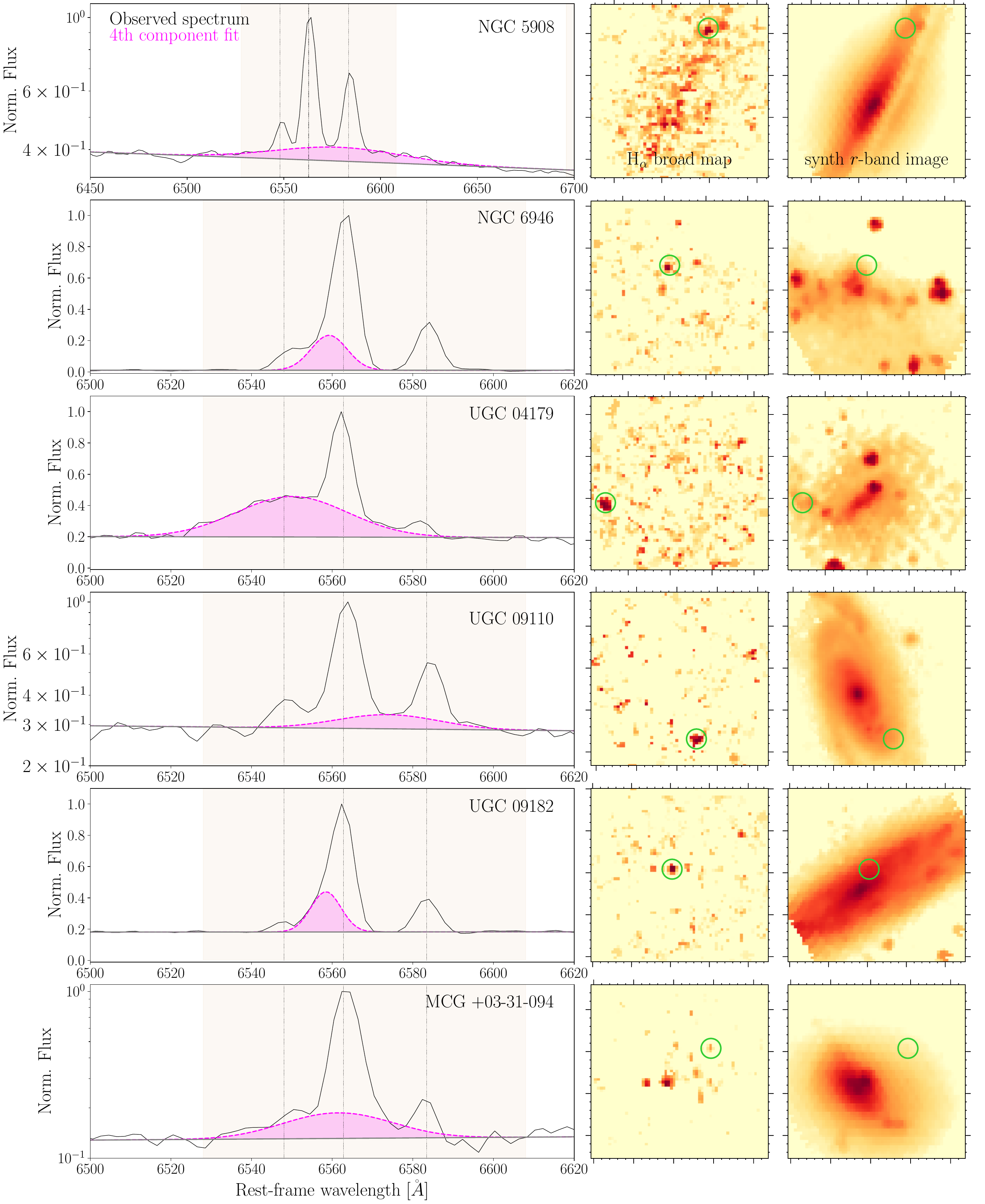}
\caption{Same as Figure~\ref{fig:group1_all_1}.}
\label{fig:group1_all_3}
\end{figure*}

%We experimented with a diagnostic based on a broad component to [\ion{O}{iii}] $\lambda$5007, but we found the results to be \redpen{what was the problem, exactly? results unclear? noisy? If you found any regions with clearly broad OIII, even if there was only one, that would be interesting to know.} 

%\redpen{We should say why we don't use the width of the [OIII] line, it is pretty extreme in Cas a and NGC4449. In Danny's 2012 paper, most objects have broad Halpha, only a few have broad OIII. I can write some text to support this if I know why you chose these criteria. Now that i read this again, did you even fit as broad OIII component, or do the F and sigma listed here correspond to narrow line emission from the galaxy?}. \lcom{hector had other criteria before, but we discussed and decided to simplify and reduce the number. Anyway, the width of OIII gave no clue}

We show the values for these two diagnostic ratios in all the spaxels of the MUSE data cube for NGC 2906 in Figure \ref{fig:impactfactor}. This plot illustrates the discriminating power of our chosen diagnostic ratios, with the spaxels that cover the site of SN~2005ip clearly deviating from the distribution of values measured in the rest of the host galaxy along both axes.

%For illustration purposes, $\log \left(\sigma_{\rm H\alpha, SNR} / \Delta \sigma_{\rm H\alpha, SNR}\right)$ is shown as a function of $\log \sigma_{\rm H\alpha, SNR}$, while the two flux to flux uncertainty ratios are shown plotted against each other. 
%\redpen{This plot does not show the two Halpha diagnostic ratios, it shows the width diagnostic ratio as a function of width. We need to remake the plot.}
%The diagnostic ratios clearly differentiate the spaxels that cover the site of SN~2005ip (shown in turquoise) from the rest of the regions in NGC2906 (shown in orange), with the majority of SN~2005ip spaxels deviating more than 3$\sigma$ from the distribution of values in the rest of the galaxy.

 %We require a minimum width of 1000 km s$^{-1}$ for this component, but we impose no lower limit on the amplitude, so the procedure only returns nonzero amplitude when there is substantial flux that is not accounted for by the narrow lines. 

We have examined the values of these diagnostic ratios in all IFS data cubes in our sample of 887 PMAS and 342 MUSE observations of nearby galaxies, consisting of more than 35 million individual spectra.
%\redpen{Might be good to list the total number of spectra examined} \textcolor{green}{done}
To minimize false detections, we restrict our search to spaxels where the signal-to-noise ratio (S/N) for $\rm{H\alpha, SNR}$ 
%and [\ion{O}{iii}] 
is higher than 5 in the residual spectra. We have set a conservative threshold of 3$\sigma$ above the median of the distribution in both diagnostic ratios to flag regions of interest as a young SNR candidates.  

As a side note, we attempted a similar search centered in the [\ion{O}{iii}] $\lambda$5007 line, but we failed to find any clear candidates. This indicates that objects with strong, broad [\ion{O}{iii}] emission like Cas A and SNR NGC 4449 (Figure~\ref{fig:YoungSNRs}) must be rare, or short-lived, or both.

\subsection{Young SNR Candidates} %1

%\textcolor{green}{20 objetos de $\sim$1000, cuantas en SFR galaxies? Add plot, cumulative total SFR, and add 20 points of the 20 galaxies with SNR. do they tend to have high SFR?}

%\textcolor{green}{this has changed a lot since the verision we first circulated. Now I do not understand how we go from 50 to 20...}

\begin{figure*}[!t]
\centering
\includegraphics[width=\textwidth]{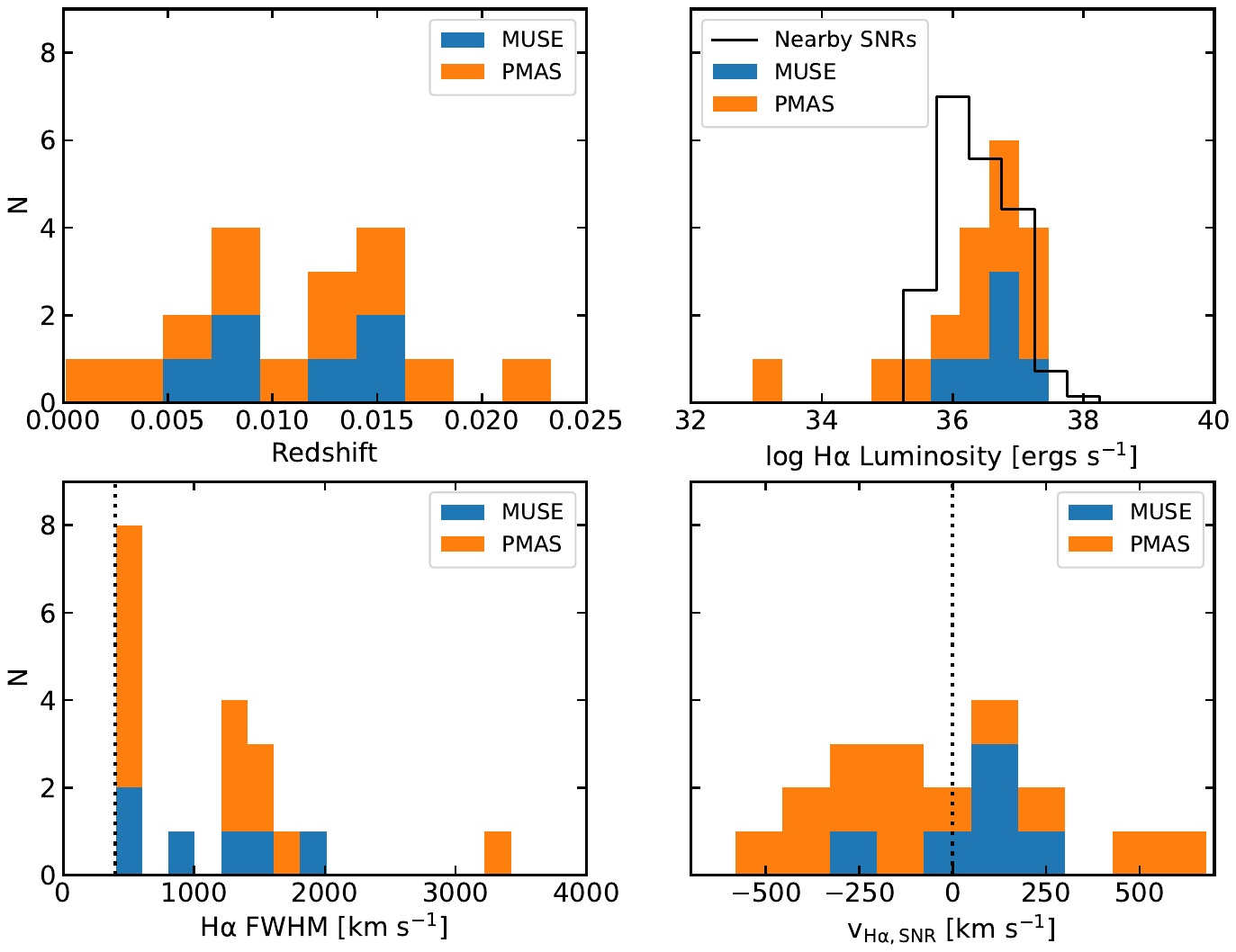}
\caption{Histograms of host redshift (top left), \ha\ luminosity (top right), \ha\ FWHM (bottom left), and \ha\ systemic velocity (bottom right) for the young SNR candiates in our sample. The histogram of \ha\ luminosities in the sample of 143 SNRs in nearby galaxies compiled by \cite{deGrijs2000} is superimposed on the top left panel, scaled down by a factor 7 for comparison. The lower limit to the \ha\ FWHM in our fits (400 \kms) is indicated by a vertical dotted line in the bottom left panel. }
\label{fig:SNRshist}
\end{figure*}

Our search yielded 20 contiguous regions that pass our quality cuts, 7 in MUSE and 13 in PMAS. We list these regions in Table~\ref{table:group1_broadHa_fluxes}, and classify them as young SNR candidates, by analogy with the properties of SN~2005ip described in Section~\ref{subsec:broad_Ha}. The only object that appears in both MUSE and PMAS data is SN~2005ip itself, which brings the total number of unique objects identified in our search to 19. The individual fits to the residual spectra around \ha\ in the spaxels identified as young SNR candidates are shown in Figures~\ref{fig:group1_all_1}, \ref{fig:group1_all_2}, and \ref{fig:group1_all_3}, along with maps of the entire host galaxy, both in the broad \ha\ component and in white light. The regions outside the SNR candidates that appear in some of these broad \ha\ maps did not pass our quality cuts. In Figure~\ref{fig:SNRshist} we show the distribution of host galaxy redshifts, along with the FWHM, luminosity, and systemic velocity of the broad \ha\ component corresponding to these regions.

%\textcolor{red}{There are actually other blobs in the broad Ha images which are not identified as SNe. Need to describe this in the text and why they are not SNe.}

Four of these 19 objects coincide spatially with previously known SNe: the already discussed SN~2005ip, a Type IIn SN in the spiral NGC 2906 imaged by PMAS and MUSE 2593 and 3104 days after discovery, respectively; ASAS-SN 14fd \citep{Holoein2019}, a Type IIn SN in the dwarf irregular galaxy LEDA 43070 (PGC 43070) imaged by MUSE 514 days after discovery; SN2011fh \citep{Pessi2022}, a Type IIn SN in the spiral NGC 4806 imaged by MUSE 1362 days after discovery; and ASAS-SN 14jb \citep{Holoein2019}, a Type IIP SN in the spiral ESO 467-51, imaged by MUSE 391 days after discovery. A fifth object, the SNR candidate in NGC 5908, coincides spatially with PSNJ15164204+5525011, a 'SN impostor' reported in 2012, likely a luminous blue variable \citep{Benetti2012}.

It is worth noting that three out of four SNe that coincide with our SNR candidates are Type IIn, despite the fact that this subtype only accounts for $\sim$9\% of CC SNe \citep{Smith2011,Kiewe2012}. 
%\textcolor{red}{HK: On the contrary, it is very much expected...JA: I don't think this is remarkable. Your technique is designed to find things that are similar to 05ip - a long-lived SNIIn, thus it is not surprising that you find a number of other long-lasting SNeIIn} 
The fourth SN, ASAS-SN 14jb, is a rare extraplanar Type IIP SN in an edge-on spiral, whose MUSE observations were analyzed and discussed in \cite{Meza2019}. The remaining 15 candidates in our sample are probably young SNRs whose SN either exploded before the era of modern transient surveys, or were missed, perhaps because of weather, or poor sampling, or coincidence with the Sun. 
%Regardless of the motive, the young SNR candidates we find do pin down the locations of the SNe that originated them. 
With distances up to $\sim$100 Mpc (for SNR Arp 142, at a redshift of 0.0233, see Figure~\ref{fig:SNRshist}), these are among the furthest SNRs identified as such. Interestingly, one of our candidate SNRs is located in NGC 6946, the `Fireworks Galaxy', a nearby spiral with a high star formation rate that has hosted 10 known SNe \citep{Eldridge2019,Eibensteiner2022}. Our results bring the total number of SNe in this galaxy up to eleven. 
%\textcolor{red}{Maybe remove this. The number of *known* SNe is unchanged -- if you include remnants then there are countless of them...}
%\textcolor{red}{In Figure 9, Why not show de Grijs as a histogram also?
%}

With all the caveats attached to small number statistics, the four objects associated with known SNe in our sample suggest that our method is turning up SNRs whose \ha\ luminosity is driven predominantly by CSM interaction, likely core collapse SNe whose progenitors have lost a great deal of mass in their pre-SN evolution, either due to winds or to binary interactions \citep{Sm09,Langer2012,Dessart2023}. Some (possibly most) of these SNe might have shown signs of interaction during their optically thick phase, appearing as Type IIn or Type Ibn SNe \citep{Kiewe2012,Taddia2013,Smith2017hsn..book..403S}. Others, like ASAS-SN 14jb, might not have developed those signs until later on, particularly if the progenitor drove some sort of fast outflow clearing a low-density cavity surrounded by denser and slower material \citep{Dwarkadas2005,Dwarkadas2007,Pat15,Pat17}. There is of course no way to tell how old these SNRs might be, but given the properties of Cas A, SNR NGC 4449, and SN2005~ip discussed in Section \ref{subsec:broad_Ha}, ages ranging between a few years and a few centuries seem reasonable.

%Although all the young SNR candidates listed in Table~\ref{table:group1_broadHa_fluxes} pass our conservative quality cuts, not all are detected at the same level of significance. For example,
Eight of our young SNR candidates have FWHM values that are within 25\% of our lower threshold of 400 \kms: 2MASX J23331223-6034201, LEDA 1015413, Arp 142, NGC 2276B, NGC 5735, NGC 6946, and UGC 09182. While these candidates might be considered somewhat more marginal than the others, it is important to emphasize that in each case the fit to the \ha\ spectral window does require the presence of a broad component with a high level of significance {\bf that shows spatial clustering in the 2D maps.} %\redpen{this is true, right?}. 
The FWHM values in the other eleven candidates range between $953\pm358$ \kms\ for NGC 4806/SN2011fh and $3423\pm713$ \kms\ for NGC 5908, comparable to the values measured in X-ray bright SNRs decades to centuries after the SN explosion \citep{Vink2012}. We note that the errors produced by MPFIT for the FWHM and systemic velocities of the broad \ha\ component are likely underestimated in the SNR candidates with the largest FWHM values. A Bayesian analysis might reveal substantial correlations in the posterior distributions for these parameters, but that is outside the scope of the present work. With one exception, the luminosities we measure for the broad \ha\ component in our SNR candidates range between $3\times10^{35}$ and $3\times10^{37}$ $\mathrm{ergs/s}$. The outlier, with a luminosity of $9\times10^{32}$ $\mathrm{ergs/s}$, is the SNR candidate in NGC 6946, which is also by far the closest galaxy in our sample \citep[$7.9\pm4.0$ Mpc,][]{Eldridge2019}. For comparison, the \ha\ luminosities of the 143 SNRs in five nearby galaxies compiled by \cite{deGrijs2000} range between $10^{36}$ and $10^{38}$ $\mathrm{ergs/s}$ (see Figure~\ref{fig:SNRshist}). Although there is considerable overlap in these luminosity ranges, it is important to keep in mind that all our SNR candidates show considerably broad emission, while most local SNRs (including those in the sample from \citealt{deGrijs2000}) do not.   

%\cite{Pessi2023}:

\begin{figure*}[!]
\centering
\includegraphics[width=0.32\textwidth]{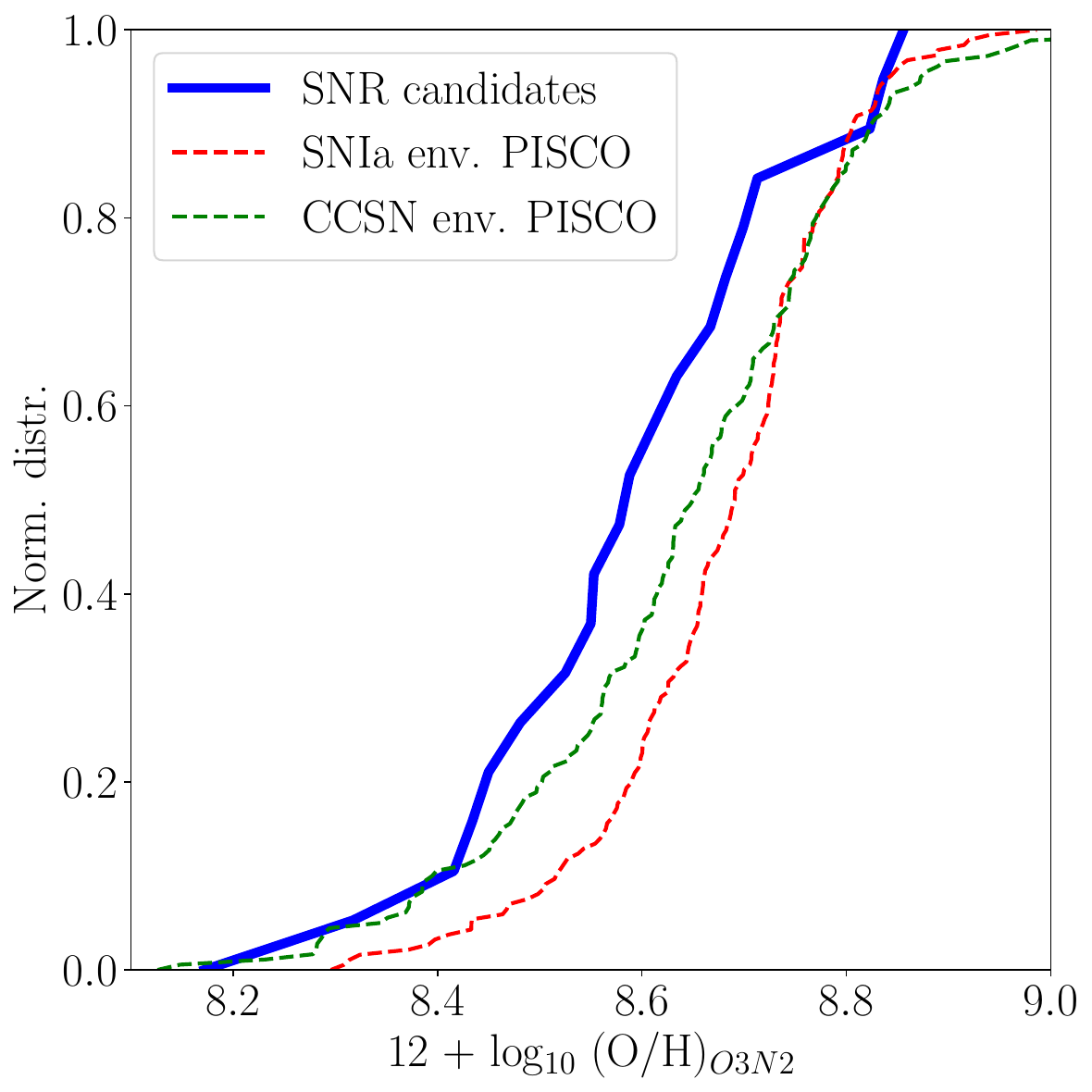}
\includegraphics[width=0.32\textwidth]{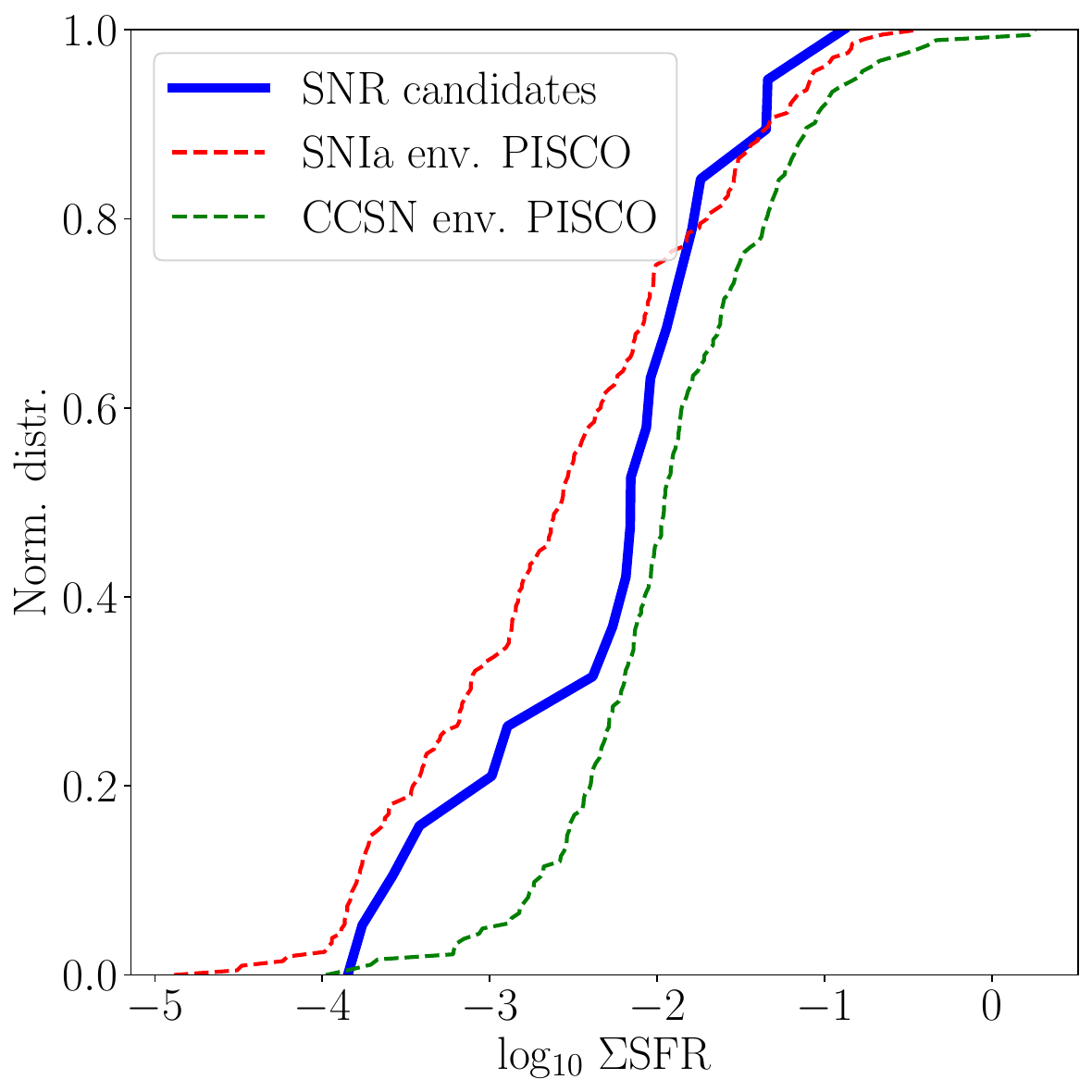}
\includegraphics[width=0.32\textwidth]{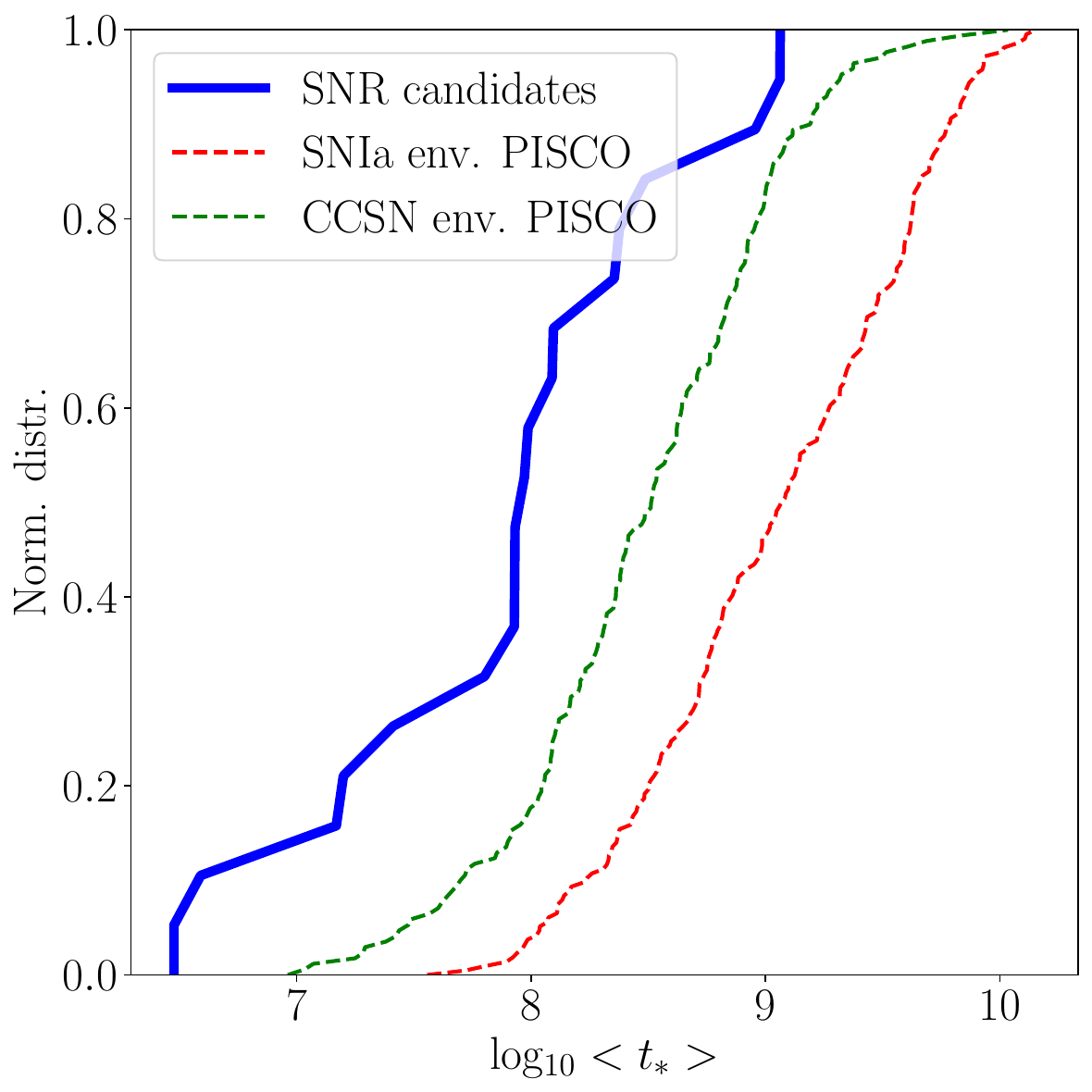}
\caption{Distribution of oxygen abundance using the O3N2 calibration from \citet{Pet04}, star-formation rate density (within 1 sq. kpc), and average stellar age for the regions of interest compared to the same parameters from all SNe Ia and CC SNe from the PISCO sample \citep{2018ApJ...855..107G}.}
\label{fig:O3N2}
\end{figure*}

In Figure~\ref{fig:O3N2} we show the distribution of the metallicities, star formation rates and average stellar ages of the candidate SNRs in our sample, derived from the IFS spectra at their locations, compared to larger samples of CC and Type Ia SNe from PISCO \citep[190 CC SNe and 234 SN Ia, respectively,][]{2018ApJ...855..107G}. 
All the parameters for our SNR candidates have been measured following the procedures described in \cite{{2018ApJ...855..107G}}. 
The metallicities at the location of the candidate SNRs are somewhat lower than those found in the environments of the PISCO CC SNe, and the star formation rates are intermediate between the CC and Ia SNe in the PISCO samples, but these differences are small and hard to interpret for a sample as small as ours. The most striking systematic difference between our candidate SNRs and the bulk population of PISCO SNe is in the average stellar ages, which are clearly lower by about half a dex than those found in the environments of CC SNe, and about a dex lower than those of SN Ia. This indicates that the progenitors of our candidate SNRs might be shorter lived, and hence more massive, than those of a typical CC SNe. A similar trend has been found for the environments of Type IIn SNe by \cite{Moriya2023}.

\section{Discussion and Conclusions}

\label{sec:conclusions}

We have conducted a systematic search for regions with broad ($\geq400$ \kms) \ha\ emission in IFS data cubes of 1229 nearby galaxies imaged by the PMAS and MUSE instruments. We have identified 19 such regions, which we classify as SNR candidates by analogy with the properties of known objects like Cas A and SNR NGC 4449-1. Indeed, four of the regions we have found coincide with the sites of previously known CC SNe, one Type IIP and three Type IIn, including the well known interacting object SN 2005ip. These coincidences, and the physical properties of the SNR candidates we have identified, suggest that the broad \ha\ emission in these regions is produced by a strong interaction between SN ejecta and some sort of dense surrounding medium. This medium could be material lost by the SN progenitor before the explosion due to stellar winds or binary interactions, which seems to be a common feature in Type IIn SNe, or a dense component of the interstellar medium associated with the formation site of the SN progenitors. The stellar ages measured from the IFS data at the sites of our SNR candidates are younger by about 0.5 dex than the ages measured at the sites of Type II SNe in the PISCO survey, indicating that the progenitors of our SNR candidates might be more massive than those of average CC SNe. 

%\textcolor{red}{This is not easy to explain... but maybe should not be the case. How come we preferentially detect more massive progenitors as young SNR, I wonder. Not an explanation, but maybe can be tied to this https://ui.adsabs.harvard.edu/abs/2023arXiv230609647M/abstract} 

The methods presented in this paper open a new window for the study of young SNRs in nearby galaxies. Systematic searches for broad line emission in IFS data cubes in current and future surveys should yield more SNR candidates like the ones we present here, and allow us to study their progenitor population in greater detail.

\acknowledgments

We acknowledge useful discussions with Sedona Price and Evan Schneider. H.M.-R. and C.B. are funded by the NASA ADAP grant NNX15AM03G S01. H.M.-R. also acknowledges support from a PITT PACC, a Zaccheus Daniel and a Kenneth P. Dietrich School of Arts \& Sciences Predoctoral Fellowship.
L.G. acknowledges financial support from the Spanish Ministerio de Ciencia e Innovaci\'on (MCIN), the Agencia Estatal de Investigaci\'on (AEI) 10.13039/501100011033, and the European Social Fund (ESF) "Investing in your future" under the 2019 Ram\'on y Cajal program RYC2019-027683-I and the PID2020-115253GA-I00 HOSTFLOWS project, from Centro Superior de Investigaciones Cient\'ificas (CSIC) under the PIE project 20215AT016, and the program Unidad de Excelencia Mar\'ia de Maeztu CEX2020-001058-M.
JDL acknowledges support from a UK Research and Innovation Future Leaders Fellowship (MR/T020784/1).
This work was funded by ANID, Millennium Science Initiative, ICN12\_009
I.D. is supported by the project PID2021-123110NB-I00 financed by MCIN/AEI /10.13039/501100011033 / FEDER, UE.
Based on observations collected at the Centro Astron\'omico Hispano en Andaluc\'ia (CAHA) at Calar Alto, operated jointly by Junta de Andaluc\'ia and Consejo Superior de Investigaciones Cient\'ificas (IAA-CSIC).
Based on observations made with ESO Telescopes at the Cerro Paranal Observatory under programs ID 095.D-0091(A), 095.D-0091(B), 096.D-0296(A), 097.D-0408(A), 098.D-0115(A), 099.D-0022(A), 0100.D-0341(A), 0101.D-0748(A), 0101.D-0748(B), 0102.D-0095(A), 0103.D-0440(A), 0104.D-0498(A), and 0104.D-0498(B).

This research was supported in part by the University of Pittsburgh Center for Research Computing, RRID:SCR 022735, through the resources provided. Specifically, this work used the H2P cluster, which is supported by NSF award number OAC-2117681.

%\hcom{Lluís, hay que poner en agradecimientos al UPitt Supercomputing Center...Carles sabe}

\vspace{5mm}
\facilities{VLT:Yepun,CAO:3.5m}
%\facility{}
%\software{astropy \citep{2013A&A...558A..33A},  
%          Cloudy \citep{2013RMxAA..49..137F}, 
%          SExtractor \citep{1996A&AS..117..393B}
%          }

\bibliography{Paper_remnants}{}
\bibliographystyle{aasjournal}

%\appendix

%\section{Outlier tables}
%\input{Table_final.tex}
%\newpage

\end{document}